\documentclass[a4paper]{article}
\usepackage{amsfonts,amssymb,amsmath,amsthm,cite}
\usepackage{ucs} 
\usepackage[utf8x]{inputenc}
\usepackage{graphicx}
\usepackage[english]{babel}
\usepackage{slashed}
\usepackage{textcomp}
\usepackage{bm}
\usepackage{titlesec}
\usepackage{stackengine}

\usepackage{etoolbox}
\patchcmd{\thebibliography}{\section*}{\section}{}{}
\usepackage{titlesec}
\usepackage{adjustbox}
\usepackage{nccmath}
\usepackage{mathtools}
\usepackage{xcolor}
\usepackage{footnote}
\usepackage{hyperref}
\usepackage{calligra} 
\hypersetup{
	colorlinks,
	citecolor=red,
	filecolor=black,
	linkcolor=red,
	urlcolor=black
}
\makesavenoteenv{tabular}
\mathtoolsset{showonlyrefs}
\evensidemargin=\oddsidemargin
\begin{document}

\vspace{10mm}
	\begin{center}
		\large{\textbf{On background fields and a cutoff in sigma models}}
\end{center}
\vspace{5mm}
\begin{center}
		\textbf{N. V. Kharuk}
	\end{center}
	\begin{center}
		St. Petersburg Department of Steklov Mathematical Institute of Russian Academy of Sciences,\\ 
		27 Fontanka, St. Petersburg 191023, Russia
	\end{center}
\begin{center}
	E-mail: natakharuk@mail.ru
\end{center}

\vspace{5mm}
	
\noindent\textbf{Abstract.} In this paper, using the example of a two-dimensional nonlinear sigma model with the Heisenberg group, we compare two variants of chiral field decomposition into a background part and a fluctuation. It is shown that only one of these methods is consistent with the construction of the generating functional by introducing a background field. Furthermore, we perform a one-loop renormalization of the quantum action, calculate power-law singularities in the two-loop approximation, and consider transition to an extended classical action. Finally, we study the consistency of the cutoff with special functional relations within the framework of the background field method.

\vspace{5mm}
\noindent\textbf{Key words and phrases:} chiral field, sigma model, Heisenberg group, background field, regularization, cutoff, averaging, renormalization, quantum action, singularity, divergence.

\vspace{20mm}

\tableofcontents
\newpage

\section{Introduction}
\label{sec-1}
Consider the standard Euclidean space $\mathbb{R}^2$, a compact semisimple Lie group $\mathrm{G}$, and its Lie algebra $\mathfrak{g}$, see \cite{2}. Consider the field $C_\mu(x)=\mathrm{U}^{-1}(x)\partial_{x^\mu}\mathrm{U}(x)$, where $\mu=1,2$ and $\mathrm{U}\in C(\mathbb{R}^2,\mathrm{G})$. It is an element of the vector space $\mathfrak{g}$ and can be decomposed into generators $t^a$ in the form $C_\mu^{\phantom{1}}(x)=C_\mu^{a}(x)t^a$ , where $a\in\{1,\ldots,\dim\mathfrak{g}\}$. It is known that the generators form a basis in $\mathfrak{g}$, are independent of the variable $x$, and are orthonormal with respect to a bilinear form $(t^a,t^b)=-\delta^{ab}$. Note that the coefficients $C_\mu^{a}(x)$ are real. In this case, the classical action for the principal chiral field model, see \cite{sig1}, is given by the formula
\begin{equation}\label{nn-1}
\mathrm{S[U]}=\frac{1}{2\gamma^2}\int_{\mathbb{R}^2}\mathrm{d}^2x\,
C_\mu^{a}(x)C_\mu^{a}(x),
\end{equation}
where $\gamma$ is a coupling constant. Such a model plays an important role in modern science, see for reference \cite{q-15,q-12,q-11,q-10,q-7}, in particular, when studying some issues in the theory of regularization and renormalization, see for example \cite{sk-b-19,Ivanov-Akac,AIK-25,i-2626}. 

In this paper, instead of $\mathrm{G}$, we consider the Heisenberg group $\mathrm{H}_3(\mathbb{R})\equiv\mathrm{H}_3$, see \cite{nn-1}. Such a group is not compact, so the classical action corresponds to a nonlinear sigma model from a broader class. On the other hand, $\mathrm{H}_3$ is nilpotent, so some issues related to the renormalization \cite{6,7,105} and the background field method \cite{102,24,25,26,sk-b-19} become more transparent. The corresponding Lie algebra $\mathfrak{h}$ is constructed using three generators
\begin{equation*}
t^1=\scriptsize\begin{pmatrix} 
	0 & 1 & 0 \\ 
	0 & 0 & 0 \\ 
	0 & 0 & 0  
\end{pmatrix}\normalsize,\,\,\,
t^2=\scriptsize\begin{pmatrix} 
	0 & 0 & 1 \\ 
	0 & 0 & 0 \\ 
	0 & 0 & 0  
\end{pmatrix}\normalsize,\,\,\,
t^3=\scriptsize\begin{pmatrix} 
	0 & 0 & 0 \\ 
	0 & 0 & 1 \\ 
	0 & 0 & 0  
\end{pmatrix}\normalsize,
\end{equation*}
with the standard matrix product. At the same time $\mathrm{H}_3=\{\exp(c_at^a):c_a\in\mathbb{R}\}$.

This paper is devoted to the discussion of two points. First, in Section \ref{sec-2}, we study variants of  decomposition of $C_\mu^a(x)$ into a background field and a fluctuation. This question is important because making the wrong choice can significantly distort and complicate the renormalization procedure. In this regard, a critical remark is made regarding the representations used in \cite{sk-b-19,Ivanov-Akac,AIK-25}. Secondly, in Section \ref{sec-3}, we write out a quantum action, calculate the first approximation, study two-loop power singularities, and discuss the renormalization process as a whole. And although similar calculations have already appeared in the case of dimensional regularization, see \cite{ee-1,ee-2}, the calculations are new for the model with a cutoff, including the averaging method. Moreover, we note that previously Laplace operators with variable principal coefficients were not studied in the context of regularization by cutoff in the coordinate representation.

\section{On background fields}
\label{sec-2}
Let us compare two variants of decomposition of the coefficients $C_\mu^a$ into a background field and a fluctuation. The first approach is a direct continuation of the idea, successfully used in scalar models \cite{Iv-2024-1,Kh-25} and Yang--Mills theory \cite{ya-21,23}, in the case of nonlinear sigma models \cite{n-11,n-12}. In this method, it is proposed to introduce a background field $b_a$ and a fluctuation $\phi_a$ on "equal rights" as a linear sum. At the same time, taking into account the one-to-one relation between $\exp(u_at^a)$ and $u^at^a$ for the model under consideration, we decompose elements of the group
\begin{equation}\label{n-12}
\mathrm{H}_3\ni \mathrm{U}=\scriptsize\begin{pmatrix} 
	1 & u_1 &u_2 \\ 
	0 & 1 & u_3 \\ 
	0 & 0 & 1  
\end{pmatrix}\normalsize=\mathrm{1}+b_at^a+\gamma\phi_at^a,\,\,\,
\mbox{where}\,\,\,u_a=b_a+\gamma\phi_a.
\end{equation}
Then we obtain
\begin{equation}\label{n-1}
\mathrm{C_\mu=U^{-1}\partial_\mu U=}\scriptsize\begin{pmatrix} 
	0 & \partial_\mu u_1 & \partial_\mu u_2-u_1\partial_\mu u_3 \\ 
	0 & 0 & \partial_\mu u_3 \\ 
	0 & 0 & 0  
\end{pmatrix}\normalsize.
\end{equation}
When implementing such a replacement, the functional $\mathrm{S}$ is a finite sum of terms $\Gamma_i[b,\phi]$, where the subscript indicates the degree of the fluctuation. At the same time, the relations are fulfilled
\begin{equation*}
\delta_{\phi_a(x)}\Gamma_i[b,\phi]=\delta_{b_a(x)}\Gamma_{i-1}[b,\phi],
\,\,\,\mbox{where}\,\,\,i\in\{1,2,3,4\}.
\end{equation*}
They are key to constructing generating functionals within the framework of the background field method, since in the diagrammatic language they can be reformulated as follows: differentiation with respect to the background field is equivalent to adding an additional tail while maintaining the existing connectivity. It is known that this procedure is valid for constructing both connected and strongly connected diagrams, see for reference \cite{Vas-98} and Sections 4.1 and 4.2 in \cite{n-32}.

The second approach is an analogue of the decomposition used in \cite{sk-b-19,Ivanov-Akac,AIK-25,i-2626}. It reduces to a nonlinear substitution of the form $\mathrm{U}=g\exp(\gamma\varphi_at^a)$, where $g$ is a matrix containing only background fields. Thus, we get
\begin{equation*}
\mathrm{U=}\scriptsize\begin{pmatrix} 
	1 & g_1 & g_2 \\ 
	0 & 1 & g_3 \\ 
	0 & 0 & 1  
\end{pmatrix}\normalsize
\scriptsize\begin{pmatrix} 
	1 & \gamma\varphi_1 & \gamma\varphi_2+\gamma^2\varphi_1\varphi_3/2 \\ 
	0 & 1 & \gamma\varphi_2 \\ 
	0 & 0 & 1  
\end{pmatrix}\normalsize=\mathrm{1}+g_at^a+\gamma\varphi_at^a+(\gamma\varphi_1/2+g_1)\gamma\varphi_3t^2,
\end{equation*}
\begin{equation*}
	\mathrm{C_\mu=U^{-1}\partial_\mu U=}\scriptsize\begin{pmatrix} 
		0 & \partial_\mu v_1 & \partial_\mu v_2-v_1\partial_\mu v_3 \\ 
		0 & 0 & \partial_\mu v_3 \\ 
		0 & 0 & 0  
	\end{pmatrix}\normalsize+\partial_\mu (\gamma\varphi_1/2+g_1)\gamma\varphi_3t^2,\,\,\,
\mbox{where}\,\,\,v_a=g_a+\gamma\varphi_a.
\end{equation*}
When comparing the two decomposition methods, it is found that in the second case an additional term arises in which the elements of the background field and fluctuations enter in different degrees. This circumstance becomes a serious obstacle in the construction of generating functionals, since differentiation by the background field cannot reproduce all available densities. For example, in this decomposition, the classical action contains a term whose functional derivative with respect to $\varphi_3(x)$ has the form
\begin{equation*}
\delta_{\varphi_3(x)}\int_{\mathbb{R}^2}\mathrm{d}^2y\,\varphi_1(y)\varphi_2(y)\varphi_3(y)=\varphi_1(x)\varphi_2(x)\neq0=
\delta_{g_i(x)}\Big(\mbox{absent}\Big).
\end{equation*}
Such a contribution cannot be reproduced by any suitable differentiation based on the background field, since there is no such term. From this point of view, the arguments concerning the choice of the background field and the construction of generating functional in the works of \cite{sk-b-19,Ivanov-Akac,AIK-25,i-2626} remain unclear. Unfortunately, the authors did not discuss this issue and did not solve the problem of the possibility of recalculation.

\section{Renormalization issues}
\label{sec-3}
\subsection{Quantum action}
\label{sec-31}

In this case, the classical action corresponds to the Euclidean version of the bosonic sigma model in a two-dimensional flat space, see \cite{ee-3} or Section 2 in \cite{n-11}. In addition, it also resembles the functional for a scalar quartic model with three fields, see \cite{29-3}. Let us rewrite the action in new terms. Next, let Planck's constant $\hbar$ play the role of a small parameter, and the new coupling constant $g$ takes finite fixed values. In this case, $\gamma=g\sqrt{\hbar}$. Then, by scaling $b_a\to gb_a$, the action can be reduced to a more familiar form. Let us denote $\mathrm{B}=\mathrm{1}+gb_at^a$, $\phi=\phi_at^a$, and $u_a=b_a+\sqrt{\hbar}\phi_a$, as well as define a new classical action using the formula
\begin{align}\nonumber
\mathrm{S_{cl}}[b+\sqrt{\hbar}\phi]=\hbar\mathrm{S}[\mathrm{B}+g\sqrt{\hbar}\phi]
&=\frac{1}{2}\int_{\mathbb{R}^2}\mathrm{d}^2x\,\Big(
(\partial_\mu u_a)(\partial_\mu u_a)-2g
u_1(\partial_\mu u_2)(\partial_\mu u_3)+
g^2 u_1^2(\partial_\mu u_3)(\partial_\mu u_3)
\Big)\\\label{eee-1}&
=\frac{1}{2}\int_{\mathbb{R}^2}\mathrm{d}^2x\,\Big(
\mathrm{w}^{ab}(u)(\partial_\mu u_a)(\partial_\mu u_b)
\Big),
\end{align}
where,in the last transition, we have used the definition
\begin{equation*}
\mathrm{w}(u)=\scriptsize\begin{pmatrix} 
		1 & 0 & 0 \\ 
		0 & 1 & -gu_1 \\ 
		0 & -gu_1 & 1+g^2u_1^2
	\end{pmatrix}\normalsize.
\end{equation*}
Also we use the notation $\mathrm{w}\equiv\mathrm{w}(gb)$. From the last representation of \eqref{eee-1}, it can be seen that the matrix $3\times3$ can be identified with the metric of the target space. Moreover, if we calculate the curvature tensor $R_{abcd}$ using this metric, then we can make sure that it is not covariantly constant, that is, $R_{abcd;e}\neq0$. This means that the action \eqref{eee-1} belongs to a more general class than the one discussed in \cite{n-11}. 

Thus, when working with a curved space, the fluctuation $\phi_a$ should be decomposed into geodesics, as was done in Section 4 of \cite{n-11}. Nevertheless, further we understand the set $\{\phi_a(x)\}_{a=1}^3$ as an element of the flat space $\mathbb{R}^3$ for each $x\in\mathbb{R}^2$. This choice is related to the peculiarity of the group $\mathrm{H}_3(\mathbb{R})$, see formula \eqref{n-12}. Indeed, by changing the values of the field $\phi_a$ with a fixed background $b_a$, we can get all the elements of the group. The following decomposition is valid for the new action
\begin{equation*}
\mathrm{S}_{\mathrm{cl}}[b+\sqrt{\hbar}\phi]=
\mathrm{S}_{\mathrm{cl}}[b]+\frac{\Gamma_1[\phi]}{g\sqrt{\hbar}}+
\frac{1}{2}\int_{\mathbb{R}^2}\mathrm{d}^2x\,\phi^a(x)\mathrm{A}^{ab}_2(x)\phi^b(x)+
g\sqrt{\hbar}\Gamma_3[\phi]+\frac{g^2\hbar}{2}\Gamma_4[\phi].
\end{equation*}
Here, the auxiliary functionals $\Gamma_i$ are proportional to the $i$-th degree of the fluctuation $\phi_a$. In this case, the explicit form of $\Gamma_1$ is not important, since it does not appear in further calculations, while $\Gamma_3$ and $\Gamma_4$ are determined by the equalities
\begin{align*}
\Gamma_3[\phi]&=\int_{\mathbb{R}^2}\mathrm{d}^2x\,
\Big(b_1\partial_{\mu}\phi_3+\phi_1\partial_\mu b_3-\partial_{\mu}\phi_2\Big)\phi_1\partial_{\mu}\phi_3,\\
\Gamma_4[\phi]&=\int_{\mathbb{R}^2}\mathrm{d}^2x\,\Big(\phi_1\partial_{\mu}\phi_3\Big)\Big(\phi_1\partial_{\mu}\phi_3\Big).
\end{align*}
The operator in the quadratic form is given by the relation $\mathrm{A}_2=\mathrm{A}_1+\mathrm{p}_\mu\partial_\mu+\mathrm{q}$, where $\mathrm{A}_1=-\partial_\mu\mathrm{w}\,\partial_\mu$, as well as
\begin{equation*}
\frac{\mathrm{p}_\mu}{g}=\scriptsize\begin{pmatrix} 
	0 & -\partial_\mu b_3 & 2gb_1\partial_\mu b_3-\partial_\mu b_2 \\ 
	\partial_\mu b_3 & 0 & 0 \\ 
	\partial_\mu b_2-2gb_1\partial_\mu b_3 & 0 & 0
\end{pmatrix}\normalsize,\,\,\,
\frac{\mathrm{q}}{g}=\scriptsize\begin{pmatrix} 
	g(\partial_\mu b_3)(\partial_\mu b_3) & 0 & 0 \\ 
	-(\mathrm{A}b_3) & 0 & 0 \\ 
	2g(\mathrm{A}b_3)b_1-2g(\partial_\mu b_3)(\partial_\mu b_1)-(Ab_2) & 0 & 0
\end{pmatrix}\normalsize.
\end{equation*}
Here, $\mathrm{A}=-\partial_\mu\partial_\mu$. It is possible to check that $\mathrm{A}_1$ and $\mathrm{A}_2$ are symmetric. The symbols $\mathrm{R}_1(\cdot,\cdot)$ and $\mathrm{R}_2(\cdot,\cdot)$ denote the corresponding fundamental solutions. By construction, they satisfy the equalities
\begin{equation}\label{n-2}
\mathrm{A}^{ac}_i(x)\mathrm{R}^{cb}_i(x,y)=\delta^{ab}\delta(x-y)
\end{equation}
in the sense of generalized functions on the Schwartz class, see \cite{Vladimirov-2002}. It is assumed that the function $\mathrm{R}_1$ is symmetric and can be uniquely fixed by suitable boundary conditions. At the same time, near the diagonal, when $x\sim y$, it has the standard behavior of the form $\mathrm{R}(x-y)\mathrm{w}^{-1}(y)$, where $\mathrm{R}(x)=-\ln(|x|\mu)/(2\pi)$, and $\mu>0$ is an auxiliary fixed parameter to make combinations dimensionless, see for reference \cite{29,30-1-1 }. The second function $\mathrm{R}_1$ is defined by the perturbative decomposition
\begin{align}\label{n-13}
\mathrm{R}_2(x,y)=\mathrm{R}_1(x,y)+
\sum_{k=1}^{+\infty}(-1)^k
\int_{\mathbb{R}^2}\mathrm{d}^2x_1\ldots\int_{\mathbb{R}^2}\mathrm{d}^2x_k\,
\mathrm{R}_1(x,x_1)\Big(\mathrm{p}_\mu&(x_1)\partial_{x_1^\mu}+\mathrm{q}(x_1)\Big)\mathrm{R}_1(x_1,x_2)\times\\\noindent\ldots&\times
\Big(\mathrm{p}_\mu(x_k)\partial_{x_k^\mu}+\mathrm{q}(x_k)\Big)\mathrm{R}_1(x_k,y).
\end{align}
By construction it is also symmetrical. Since $\mathrm{R}_i$ are singular at $x\sim y$, for further study, we introduce a regularization by deforming $\mathrm{R}_1^{\phantom{1}}\to\mathrm{R}_1^\Lambda$, where the parameter $\Lambda\gg1$ is regularizing one. At the same time, the corresponding substitution of $\mathrm{R}\to\mathrm{R}^\Lambda$ takes place in the main order near the diagonal. For deformation, for example, an averaging method can be used, see \cite{Ivanov-2022,Iv-2024}. We do not choose uniquely the type of the regularization, but we require only a few standard conditions concerning the main order:
\begin{equation}\label{n-3}
\mathrm{R}^\Lambda(0)=\frac{\ln(\Lambda/\mu)}{2\pi}+\mathcal{O}(1),\,\,\,
\partial_\mu\mathrm{R}^\Lambda(0)=\mathcal{O}(1),\,\,\,\Lambda^{-2}\mathrm{A}\mathrm{R}^\Lambda(0)=\mathcal{O}(1),
\end{equation}
where equalities should be understood in the sense of the asymptotic expansion with respect to $\Lambda$. The function $\mathrm{R}_2$ after deformation is denoted by $\mathrm{R}_2^\Lambda$. Taking into account the latest definitions, the regularized connected quantum action $\mathrm{W}_{\mathrm{reg}}^{\mathrm{c}}$ has the form of the functional integral, see \cite{Vas-98},
\begin{equation*}
\mathrm{W}_{\mathrm{reg}}^{\mathrm{c}}[b]=-\hbar\ln\bigg(\int\mathcal{D}\phi\,e^{-\mathrm{S}_{\mathrm{cl}}[b+\sqrt{\hbar}\phi]/\hbar}\bigg)\bigg|_{\mathrm{reg.}},
\end{equation*}
which must be understood in the sense of a formal series by Planck's constant
\begin{equation*}
\mathrm{W}_{\mathrm{reg}}^{\mathrm{c}}[b]=\mathrm{S}_{\mathrm{cl}}[b]-\frac{\hbar}{2}\Big(\ln\det\big(\mathrm{R}_1^\Lambda\big)-\kappa_0\Big)-\hbar
\bigg[\mathbb{H}_0^{\mathrm{c}}\exp\Big(-\Gamma_1/(g\sqrt{\hbar})-g\sqrt{\hbar}\Gamma_3-g^2\hbar\Gamma_4/2\Big)-\sum_{k=1}^{+\infty}(g^2\hbar)^k\kappa_{k}\bigg].
\end{equation*}
Here, the functionals $\Gamma_1$, $\Gamma_3$, and $\Gamma_4$ are related to vertices with one, three, and four external lines, and the operator $\mathbb{H}_0^{\mathrm{c}}$ connects all free ends in pairs by means of $\mathrm{R}_2^\Lambda$ in all possible ways, leaving only the connected part without external lines. The constants $\kappa_k$ subtract singularities independent of the background field $b_a$. At the same time, choosing the background field by solving a quantum equation of motion, only strongly connected parts remain in total, the vertex $\Gamma_1$ disappears, and the superscript in the operator and the action changes $\mathrm{c}\to\mathrm{sc}$. Since renormalization issues can be studied using the example of the strongly connected part, we continue further considerations with it, leaving the background field unfixed. The following decomposition holds
\begin{equation*}
\mathrm{W}_{\mathrm{reg}}^{\mathrm{sc}}[b]=\mathrm{S}_{\mathrm{cl}}[b]+\hbar\mathrm{W_1}[b]+g^2\hbar^2\mathrm{W_2}[b]+\mathcal{O}(\hbar^3),
\end{equation*}
where
\begin{equation}\label{ns-1}
\mathrm{W_1}[b]=-\frac{1}{2}\Big(\ln\det\big(\mathrm{R}_1^\Lambda\big)-\kappa_0\Big)
-\frac{1}{2}\ln\det\big(\mathrm{R}_2^\Lambda/\mathrm{R}_1^\Lambda\big).
\end{equation}
At the same time, in the second approximation
\begin{equation}\label{nn-2}
\mathrm{W_2}[b]=\frac{1}{2}\Big(\mathbb{H}_0^{\mathrm{sc}}\big(\Gamma_4^{\phantom{1}}\big)-\mathbb{H}_0^{\mathrm{sc}}\big(\Gamma_3^{2}\big)\Big)+\hat{\kappa}_1
\end{equation}
only a part that contains power-law ($\sim\Lambda^2$) singularities is studied. An explicit expression of the corresponding functional, taking into account counterdiagrams, is given in Section \ref{sec-st}.\\

\noindent\textbf{Remark on regularization.} The possibility of choosing the background field as a solution to the quantum equation of motion depends, in particular, on the possibility of introducing regularization in a special coordinated way, see Section 4.1 in \cite{Iv-2024-1}, so that the relationship between the density of the quantum equation of motion and action is preserved
\begin{equation*}
\frac{\delta \mathrm{W}_{\mathrm{reg}}^{\mathrm{sc}}[b]}{\delta b_a(x)}=Q^a_{\mathrm{reg}}[b](x).
\end{equation*}
In this case, the renormalization process is also consistent. Otherwise, serious problems arise related to the adjustment of the renormalization process, see a similar situation in the Yang--Mills theory \cite{n-32}. If we consider a cutoff as regularization, in particular, by means of averaging, then in the case of sigma models the main difficulty lies in the feasibility of the "chain" relation for regularized Green functions.
\begin{equation}\label{n-14}
\frac{\delta \mathrm{R}_{i}^{\Lambda ab}(x,y)}{\delta b_a(z)}=-
\mathrm{R}_{i}^{\Lambda ab}(x,z)\Big(\partial_{b_a(z)}\mathrm{A}_i(z)\Big)\mathrm{R}_{i}^{\Lambda ab}(z,y).
\end{equation}
Moreover, given the expansion from \eqref{n-13}, the validity of the condition at $i=1$ guarantees its existence at $i=2$. Let us introduce a few auxiliary objects
\begin{equation}\label{n-11}
\mathrm{w}^{-1}=\scriptsize\begin{pmatrix} 
	1 & 0 & 0 \\ 
	0 & 1+g^2b_1^2 & gb_1 \\ 
	0 & gb_1 & 1
\end{pmatrix}\normalsize,\,\,\,
\mathrm{v}=\scriptsize\begin{pmatrix} 
	-1 & 0 & 0 \\ 
	0 & 0 & 1 \\ 
	0 & 1 & -gb_1
\end{pmatrix}\normalsize,\,\,\,
\mathrm{v}^{-1}=\scriptsize\begin{pmatrix} 
	-1 & 0 & 0 \\ 
	0 & gb_1 & 1 \\ 
	0 & 1 & 0
\end{pmatrix}\normalsize.
\end{equation}
It is not difficult to verify that $\mathrm{w}=\mathrm{v}\mathrm{v}$ and $\det(\mathrm{v})=1$. This decomposition can be understood as a transition to a tetrad (vierbein) formalism. Define the operator $\mathrm{A}_0=\mathrm{v}\mathrm{A}\mathrm{v}$, where $\mathrm{A}=-\partial_\mu\partial_\mu$, and its standard Green's function denote as $\mathrm{R}_0(x,y)=\mathrm{v}^{-1}(x)\mathrm{R}(x-y)\mathrm{v}^{-1}(y)$. Next, we decompose $\mathrm{R}_{1}$ with respect to the powers of the first-order operator $\mathrm{A}_1-\mathrm{A}_0=\mathrm{v}(\mathrm{\hat{p}}_\mu\partial_\mu+\mathrm{\hat{q}})\mathrm{v}$, where
\begin{equation*}
	\frac{\mathrm{\hat{p}}_\mu}{g}=\scriptsize\begin{pmatrix} 
		0 & 0 & 0 \\ 
		0 & 0 & \partial_\mu b_1 \\ 
		0 & -\partial_\mu b_1 & 0
	\end{pmatrix}\normalsize,\,\,\,
	\frac{\mathrm{\hat{q}}}{g}=\scriptsize\begin{pmatrix} 
		0 & 0 & 0 \\ 
		0 & g(\partial_\mu b_1)(\partial_\mu b_1) & 0 \\ 
		0 & \mathrm{A}b_1 & 0
	\end{pmatrix}\normalsize,
\end{equation*}
using $\mathrm{R}_{0}$, as was done in \eqref{n-13}. Then the question of the validity of relation \eqref{n-14} for $i=1$ follows from its feasibility for $i=0$. It is at this stage that the main difficulties arise, because the direct use of regularization by a cutoff leads to a transition to the deformed function
\begin{equation}\label{n-15}
\mathrm{R}_{0}^{\phantom{1}}(x,y)\to\mathrm{R}_{0}^\Lambda(x,y)=\mathrm{v}^{-1}(x)\mathrm{R}^\Lambda(x-y)\mathrm{v}^{-1}(y).
\end{equation}
However, it does not satisfy equality \eqref{n-14}. The question of other possible deformations remains open. For example, in the case of a small background field, as a Green's function $\mathrm{R}_{0}$, we can write out a formal series by powers of the difference $\mathrm{A}_0-\mathbf{1}\cdot\mathrm{A}$. The problem is that such a difference is a second-order operator, so the question of convergence in this case also remains open. Further, we use regularization in the form of \eqref{n-15}. This is acceptable, since the main singularities are considered below, which either do not depend on the type of regularization (the first correction), or are dictated by the main approximation of the Green's function (power-law parts in the second correction).

\subsection{First correction}

Let us primarily consider the first part of \eqref{ns-1}. Given $\det(\mathrm{w})=1$, as $\kappa_0$ we choose $\ln\det(\mathrm{R}_0^\Lambda)$, which actually does not depend on the background field and is equal to $\ln\det(\mathrm{R}^\Lambda)$. Next, using the reasoning from Section \ref{sec-31}, we apply the relation $\ln\det(\cdot)=\mathrm{tr}\ln(\cdot)$, then we obtain a perturbative decomposition for the first part in the form
\begin{multline*}
	\sum_{k=1}^{+\infty}\frac{(-1)^{k-1}}{2k}
	\int_{\mathbb{R}^2}\mathrm{d}^2x_1\ldots\int_{\mathbb{R}^2}\mathrm{d}^2x_k\,\mathrm{tr}\Big(
	\big[\mathrm{\hat{p}}_\mu(x_1)\partial_{x_1^\mu}+\mathrm{\hat{q}}(x_1)\big]\mathrm{R}^\Lambda(x_1-x_2)\times\ldots\\
	\times\big[\mathrm{\hat{p}}_\mu(x_k)\partial_{x_k^\mu}+\mathrm{\hat{q}}(x_k)\big]\mathrm{R}^\Lambda(x_k-x_1)\Big),
\end{multline*}
where the operation $\mathrm{tr}$ denotes the standard matrix trace. It follows from the decomposition that in the two-dimensional case, only the first two terms can provide a singular contribution. By shifting variables, using the equalities from \eqref{n-3}, and removing the finite parts, we get
\begin{equation}\label{n-4}
\frac{L}{16\pi^2}\int_{\mathbb{R}^2}\mathrm{d}^2x\,\mathrm{tr}\Big(4\mathrm{\hat{q}}+\mathrm{\hat{p}}_\mu\mathrm{\hat{p}}_\mu\Big)=\frac{Lg^2}{8\pi}
\int_{\mathbb{R}^2}\mathrm{d}^2x\,(\partial_\mu b_1)(\partial_\mu b_1),
\end{equation}
where $L=\ln(\Lambda/\sigma)$. Here, $\sigma>0$ is an auxiliary parameter to make combinations dimensionless, the theory does not depend on it. For the second part of the quantum correction $\mathrm{W_1}[b]$, the following perturbative decomposition is valid
\begin{multline*}
	\sum_{k=1}^{+\infty}\frac{(-1)^{k-1}}{2k}
	\int_{\mathbb{R}^2}\mathrm{d}^2x_1\ldots\int_{\mathbb{R}^2}\mathrm{d}^2x_k\,\mathrm{tr}\Big(
\big[\mathrm{p}_\mu(x_1)\partial_{x_1^\mu}+\mathrm{q}(x_1)\big]\mathrm{R}_1^\Lambda(x_1,x_2)\times\ldots\\
	\times\big[\mathrm{p}_\mu(x_k)\partial_{x_k^\mu}+\mathrm{q}(x_k)\big]\mathrm{R}_1^\Lambda(x_k,x_1)\Big).
\end{multline*}
It can be seen from the construction that ultraviolet singularities can again be contained only in the first two terms. Repeating similar steps, we get
\begin{equation*}\label{n-5}
\frac{L}{16\pi}\int_{\mathbb{R}^2}\mathrm{d}^2x\,\mathrm{tr}\big(4\,\mathrm{w}^{-1}\mathrm{q}+\mathrm{w}^{-1}\mathrm{p}_\mu\mathrm{w}^{-1}\mathrm{p}_\mu\big)=
\frac{Lg^2}{8\pi}\int_{\mathbb{R}^2}\mathrm{d}^2x\,\Big((\partial_\mu b_3)(\partial_\mu b_3)-
(\partial_\mu b_2-gb_1\partial_\mu b_3)(\partial_\mu b_2-gb_1\partial_\mu b_3)\Big).
\end{equation*}
The derivation also used the fact that the trace of the product of the matrix $\mathrm{p}_\mu$ by any finite power of the matrix $\mathrm{w}$ is zero. Summarizing the results, we get that the singular part of $\hbar\mathrm{W_1}[b]$ is equal to
\begin{equation}\label{n-6}
-\frac{Lg^2\hbar}{4\pi}\bigg(\mathrm{S_{cl}}[b]-\int_{\mathbb{R}^2}\mathrm{d}^2x\,\Big[(\partial_\mu b_1)(\partial_\mu b_1)+(\partial_\mu b_3)(\partial_\mu b_3)\Big]\bigg).
\end{equation}
It is not proportional to the classical action. Therefore, the individual parts of the classical action must have their own renormalization constants. This situation fits into the general theory, for this it is necessary to introduce four renormalization constants
\begin{equation}
b_i\to b_i\sqrt{Z_i},\,\,\,\phi_i\to \phi_i\sqrt{Z_i},\,\,\,
g\to gZ_4/\sqrt{Z_1Z_2Z_3},
\end{equation}
which are series in powers of $g^2\hbar$ with singular coefficients $Z_i=1+g^2\hbar z_i+\mathcal{O}(\hbar^2)$. Then, taking into account the result of \eqref{n-6}, we get the following answers
\begin{equation}\label{n-16}
z_i=(-1)^iL/(4\pi).
\end{equation}
This result is completely consistent with the case of dimensional regularization, see \cite{ee-1,ee-2}. When working with sigma models, the metric is renormalized and the first correction leads to a shift to the Ricci tensor. Indeed, if we use the metric $\mathrm{w}$ to calculate the Ricci tensor, we get
\begin{equation}
\mathrm{Ric}=\frac{1}{2}\scriptsize\begin{pmatrix} 
		-1 & 0 & 0 \\ 
		0 & 1 & -gb_1^{\phantom{1}} \\ 
		0 & -gb_1^{\phantom{1}} & g^2b_1^2-1
	\end{pmatrix}\normalsize.
\end{equation}
Thus, a shift by the constants from \eqref{n-16} is equivalent to the shift $\mathrm{w}\to\mathrm{w}+\hbar g^2L\mathrm{Ric}/(2\pi)$.

\subsection{Power-law singularities}
\label{sec-st}
Let us consider power-law singularities in the second quantum correction. After one-loop renormalization, the two-loop functional is equal to the sum of $\mathrm{W_2}[b]$ from \eqref{nn-2} and the following set of counterdiagrams
\begin{equation*}
\mathrm{D_c}=\frac{g^2\hbar^2}{2}\mathbb{H}_0^{\mathrm{sc}}\Bigg(\int_{\mathbb{R}^2}\mathrm{d}^2x\,\bigg(z_1(\partial_\mu \phi_1)(\partial_\mu \phi_1)+2z_1(\partial_\mu \phi_3)(\partial_\mu \phi_3)+z_2\sum_{a,b=2,3}\phi_a\mathrm{A}_1^{ab}\phi_b\bigg)\Bigg).
\end{equation*}
We should immediately note that due to the second property of \eqref{n-3} and spherical symmetry, it follows that power-law singularities can only be of the form $\sim\Lambda^2$. To find them, it is necessary to track only those terms in diagrams in which all derivatives act on the singular function $\mathrm{R}^\Lambda(x-y)$ from the main order of decomposition of the Green's function
\begin{equation}\label{nn-3}
\mathrm{R}_2^{\Lambda}(x,y)=\mathrm{R}^\Lambda(x-y)\mathrm{w}^{-1}(y)+\ldots,
\end{equation}
Here, the ellipsis denotes less singular contributions that are finite before and after the introduction of the regularization for all values of the argument. The fact that $\mathrm{w}$ has a block diagonal form implies the absence of $\sim\Lambda^2$ terms in the counterdiagrams of $\mathrm{D_c}$. To study the remaining contributions, we introduce auxiliary functions
\begin{equation*}
\Big(\mathrm{R}_2^{\Lambda ab}(x,x)\Big)\Big|_{a=b=1}=\mathrm{R}^\Lambda(0)+f(x),\,\,\,
\frac{\alpha_1}{2\pi}=-\partial_\mu\partial_\mu\mathrm{R}^1(0),\,\,\,
\frac{\alpha_2}{2\pi}=\int_{\mathbb{R}^2}\mathrm{d}^2x\,\Big(\partial_\mu\partial_\mu\mathrm{R}^1(x)\Big)^2.
\end{equation*}
Then, substituting expression \eqref{nn-3}, taking into account the explicit form of the vertices, we have
\begin{align*}
\mathbb{H}_0^{\mathrm{sc}}\big(\Gamma_4^{\phantom{1}}\big)&\longrightarrow
\frac{\Lambda^2\alpha_1}{2\pi}\int_{\mathbb{R}^2}\mathrm{d}^2x\,f(x)+\big(\ln\mbox{-part}\big),
\\
\mathbb{H}_0^{\mathrm{sc}}\big(\Gamma_3^{2}\big)&\longrightarrow
\frac{\Lambda^2\alpha_2}{2\pi}\int_{\mathbb{R}^2}\mathrm{d}^2x\,f(x)+\big(\ln\mbox{-part}\big).
\end{align*}
Therefore, remembering the explicit form of the diagrams from \eqref{nn-2}, we obtain a power-law singularity of the form in the second quantum correction
\begin{equation*}
\Lambda^2g^2\hbar^2\frac{\alpha_1-\alpha_2}{4\pi}\int_{\mathbb{R}^2}\mathrm{d}^2x\,f(x).
\end{equation*}
To reduce it, a counterdiagram of the form must be added to the quantum action
\begin{equation}\label{nn-4}
	-\Lambda^2g^2\hbar\frac{\alpha_1-\alpha_2}{4\pi}\mathbb{H}_0^{\mathrm{sc}}\big(\mathrm{S}_2[\phi]\big),\,\,\,\mbox{where}\,\,\,\mathrm{S}_2[\phi]=\int_{\mathbb{R}^2}\mathrm{d}^2x\,\phi_1^2.
\end{equation}
In this case, there are two ways to understand the renormalization process: either by extending the renormalization condition, and then adding an auxiliary vertex $\mathrm{S}_2[\phi]$ is the final step, or by extending the classical action by adding a "mass" term, that is
\begin{equation*}
\mathrm{S_{cl}}[b+\sqrt{\hbar}\phi]\to\mathrm{S_{cl}}[b+\sqrt{\hbar}\phi]+\mu^2\hbar\mathrm{S}_2[\phi].
\end{equation*}
In the second case, the renormalization consists of finding for the constant $Z_\mu=1+g^2\hbar z_\mu+\mathcal{O}(\hbar^2)$. In this case, the coefficient $z_\mu$ has a power-law part $z_\mu^\Lambda$, which, taking into account \eqref{nn-4}, is equal to $-\Lambda^2(\alpha_1-\alpha_2)/(4\pi\mu^2)$, and can also have a logarithmic part $z_\mu^L$, which is determined based on the analysis of two-loop logarithmic singularities. Note that in the first correction, there are no new singularities related to the parameter $\mu^2$, which indicates consistency. Thus, in the second approach, power-law singularities lead to a shift in "mass".

\section{Conclusion}

The paper presents a comparison of two approaches to decomposing the chiral field into a background part and a fluctuation. It has been shown that the standard linear substitution is more well-justified, since it is consistent with the process of constructing generating functionals by introducing the background field. Additionally, an explicit form of the quantum action was presented and the singular contribution for the first correction was calculated. It turned out that the individual parts of the classical action have different renormalization constants. This leads to the reformulation of the model as a scalar one with three fields and one new interaction constant, in which the renormalization process has a standard form. At the same time, consistency with the renormalization of the metric, which was adopted in the framework of working with sigma models, was shown. Two-loop power-law singularities, options for extending the classical action, and issues of consistency of the regularization with special functional relations were also studied.

Note that when searching for renormalization constants for $\mathrm{S_{cl}}$, the fact that the first correction coefficients for $Z_2$ and $Z_4$ are equal to each other was used. This is necessary because the fifth term of the classical action is proportional to $g^2$. Accordingly, it is assumed that similar additional relations should be applied in higher corrections. This is necessary for the renormalization of the model. Another possibility when expanding the action is to introduce a new constant $\lambda$ instead of $g^2$. However, in this case, difficulties arise due to non-linearity, since the determinant of new matrix $\mathrm{w}$ depends on the fields, so the denominator in $\mathrm{w}^{-1}$ also contains the fields. Additionally, the situation is complicated by the presence of power-law singularities. Thus, the open question is related to the renormalizability of the model as a whole.

Another difficulty is related to the consistency of the method of background field and a cutoff in the coordinate representation, see Remark in Section \ref{sec-31}. This is due to the substitution of the fluctuation fields $\phi\to \mathrm{v}\phi$, see \eqref{n-11}. This, in particular, leads to a nonlinear combination of the background field and fluctuations (relative to each other), which eventually violates the consistency between the renormalization of the action and the density of the equation. It may seem that the problem can be solved by replacing not only the fluctuation, but also the background field before the introduction of regularization: $u\to\mathrm{v}|_{b=u}u$, where $u=b+\phi$. However, this approach does not correct the situation, since the leading order of the Laplace operator still has a coefficient that depends on the background field. Moreover, in the case of the Heisenberg group, the classical action \eqref{eee-1} visually remains the same and differs only by a permutation of fields. 

Section \ref{sec-3} discussed the problem of decompositions on a background field and a fluctuation used in \cite{sk-b-19,Ivanov-Akac,AIK-25,i-2626}. As an open problem, we can single out a description of the relationship between two approaches to introducing the background field, see Section \ref{sec-2}, as well as an explicit recalculation procedure. The main difficulty lies in the fact that in the general case there is no equality $\exp(\mathfrak{a})\exp(\psi)=\exp(\mathfrak{a}+\psi)$, where $\mathfrak{a},\psi\in\mathfrak{g}$, and $\exp(\mathfrak{a})$ describes the background field from $\mathrm{G}$.

\vspace{2mm}
\noindent\textbf{Acknowledgements.} The author thanks A.V.Ivanov for his criticism and editorial work.




\end{document}